\documentclass[aps,prb,floatfix,twocolumn]{revtex4}
\usepackage{graphicx}
\usepackage{dcolumn}
\usepackage{dcolumn}
\usepackage{amssymb}
\usepackage{amsmath}

\begin{document}

\title{Flat branches and pressure amorphization}

\author{Morrel H. Cohen, Jorge \'I\~niguez, and J. B. Neaton}

\address{Department of Physics and Astronomy, Rutgers University,
Piscataway, New Jersey 08854-8019, USA}

\begin{abstract}
  After summarizing the phenomenology of pressure amorphization (PA),
  we present a theory of PA based on the notion that one or more
  branches of the phonon spectrum soften and flatten with increasing
  pressure.  The theory expresses the anharmonic dynamics of the flat
  branches in terms of local modes, represented by lattice Wannier
  functions, which are in turn used to construct an effective
  Hamiltonian.  When the low-pressure structure becomes metastable
  with respect to the high-pressure equilibrium phase and the relevant
  branches are sufficiently flat, transformation into an amorphous
  phase is shown to be kinetically favored because of the
  exponentially large number of both amorphous phases and reaction
  pathways. In effect, the critical-size nucleus for the first-order
  phase transition is found to be reduced to a single unit cell, or
  nearly so. Random nucleation into symmetrically equivalent local
  configurations characteristic of the high-pressure structure is then
  shown to overwhelm any possible domain growth, and an ``amorphous''
  structure results.
\end{abstract}

\maketitle

\section{Introduction}

Since its first observation in Gd$_2$(M$_2$O$_4$) in 1972,\cite{brix}
and especially in ice nearly twenty years ago,\cite{mish}
amorphization induced by pressure (or ``pressure amorphization'', PA)
has been discovered in materials of essentially all binding types,
from ionic to metallic, which have more than one atom per primitive
cell. The disappearance of the X-ray diffraction pattern of the
low-pressure crystalline phase, or a concomitant broadening of peaks
in the vibrational spectra, is taken to be evidence of an emerging
amorphous state. These signatures typically appear within a pressure
window whose average value and overall range are strongly material
dependent.

The phenomenology and simulations of materials displaying pressure
amorphization have been thoroughly
reviewed.\cite{pony,scio,sharm,richet,greaves} It is not sufficiently
generally recognized within that extensive literature that there are
two distinct classes of materials which could be labelled as pressure
amorphized. In the first, a conventional glass, the atomic structure
is not topologically equivalent to any crystalline structure. In the
second, random displacements of the nuclei occur which do not destroy
crystalline topology.

Despite extensive investigation, and in part because the above
distinction has not been clearly made, a general physical picture of
the microscopic processes underlying PA has yet to emerge. On the
theoretical side, the main efforts have focused on two materials
exhibiting the second class of PA, $\alpha$-SiO$_2$
($\alpha$-quartz)\cite{Tse91, bing92, cha93, Som94, bing94, Wat95,
Tse97, baroni, wentz, dean} and $\alpha$-AlPO$_4$.\cite{Tse92, keskar,
cha93, Wat95} Both interatomic potentials and fully first-principles
approaches have been used to study these systems. While it remains
uncertain whether PA has been actually observed in a simulation, it
seems to be generally accepted that there is a correlation between
pressure amorphization and the softening (i.e., phonon frequencies
approaching zero) of a nearly dispersionless, or flat, acoustic phonon
branch. In addition, it has been suggested that the occurrence of an
amorphous phase should be strongly related to the dynamics of the
system.\cite{cha93,dean}

In this paper, we discuss an explicit model for materials exhibiting
the second class of PA. On the basis of inferences drawn from both
phenomenology and simulations, we first argue that these materials
share many common features and then show how to construct an effective
Hamiltonian that incorporates these features. As we demonstrate, this
microscopic model is simple enough to analyze in detail yet still
relevant to real materials.

The paper is organized as follows. In Section~II we discuss the
elements of our theory of PA on the basis of both phenomenology and
simulations. In Section~III we review some elements of lattice
vibration theory. In Section~IV, we present an effective Hamiltonian
suitable for the discussion of the physics underlying PA, and reduce
it to the simplest model capable of incorporating the essential
features discussed in Section~II. In Section~V, we describe how PA is
manifested in the simple model through an analysis of
the topology of the model's potential energy surfaces.  Sections~VI,
VII, and~VIII address the conditions for the occurrence of pressure
amorphization in the 2-site, 3-site, and $N$-site models respectively.
In Section~IX, we discuss the relation of our considerations to
conventional nucleation and growth models of structural
transitions. We summarize our results and draw conclusions in
Section~X.

\section{Phenomenology and simulations}

Many transition patterns have been observed during pressure
amorphization.\cite{pony,sharm,richet} Of these, we select five
(types~I-V) for explicit discussion:

Type~I materials transform from a crystalline phase $C$ to an
amorphous phase $a$ at pressure $p = p_{ca}$. Then, after $p$ is
reduced from above $p_{ca}$, they remain amorphous down to a lower
pressure $p_{ac}$, below which they revert to $C$, as is the case, for
example, in AlPO$_4$.\cite{Tse92} (The hysteresis $p_{ca}$-$p_{ac}$ is
large, typically of the order of $p_{ca}$ itself.) Materials of type
II remain amorphous upon release of pressure, making possible detailed
examination of the properties of the amorphous phase. This appears to
be the case for amorphous $\alpha$-SiO$_2$, which has been found to
retain elastic anisotropy.\cite{mcneil,tse} Another example is
crystalline Fe$_2$SiO$_4$-fayalite, which is antiferromagnetic; the
pressure-amorphized fayalite remains antiferromagnetic with the Neel
temperature essentially unchanged,\cite{richard,will,krueg} from
which we infer that the $a$- and $C$-phase topologies are the
same. Type~III materials become amorphous above $p_{ca}$, later
transforming to a new crystal structure $C'$ when $p$ is held steady
above $p_{ca}$ for sufficiently long. Materials of type~IV behave the
same as those of type~I as $p$ is reduced from above $p_{ca}$, but
they transform to a new crystal structure $C'$ from the amorphous
phase $a$ when pressure is increased beyond $p_{ac'}>p_{ca}$. Finally,
materials of type~V transform to $C'$ at $p_{ac'}$ upon decompression
from above $p_{ac}$, finally reverting to $C$ below $p_{cc'}$.

As already mentioned in the Introduction, lattice dynamical
calculations have shown a considerable softening of the entire lowest
band of $\alpha$-quartz, with a soft mode instability occurring first
at the zone boundary at pressures at which PA is observed\cite{bing94,
wentz} or at somewhat higher pressures.\cite{baroni} Also, it has been
argued that the transition mechanism in $\alpha$-AlPO$_4$ is similar
to that of $\alpha$-quartz.\cite{cha93,keskar}

From the phenomenology and calculations described above, the following
picture of PA emerges: The random displacements in the amorphous phase
are associated with the incipient instability of at least one nearly
flat phonon branch. The amorphous phase is the result of a kinetically
hindered transition between two closely-related crytalline phases $C$
and $C'$, where $C$ is a low-pressure crystalline phase and $C'$ is a
high-pressure one. The transition from $C$ to $C'$ is expected to be
strongly first order, and also displacive, meaning that the atomic
structure of $a$ remains topologically equivalent to that of
$C$. (This latter feature is to be contrasted with the significant
changes in topology that usually accompany ordinary glass formation.)
In the amorphous phase $a$, random displacements are large enough to
eliminate the diffraction pattern through reduction of the
Debye-Waller factor, but not so large so as to disrupt the topology of
the crystal structure associated with $C$.

\section{Nearly flat branches; local modes as lattice Wannier
functions}

Let us assume that we can work in the Born-Oppenheimer approximation,
supressing all the electronic degress of freedom. The Hamiltonian is,
then,
\begin{equation}
H[P,R] = T[P] + V[R],
\label{eq:hamil}
\end{equation}
where $P$ and $R$ are the sets of nuclear momenta and
positions. $T[P]$ is the nuclear kinetic energy,
\begin{equation}
T[P] = \sum_i \frac{P^2_i}{2M_i},
\label{eq:kinetic}
\end{equation}
where $P_i$ is the momentum of the $i$-th nucleus and $M_i$ is its
mass. $V[R]$ is the nuclear potential energy
\begin{equation}
V[R] = E[R] + V_{NN}[R],
\end{equation}
where $E[R]$ is the electronic ground state energy and $V_{NN}[R]$ the
internuclear interaction energy for configuration $R$.

Let $R_0$ be a set of positions corresponding to a local minimum of
the energy that we choose as a reference (e.g. the low-pressure
phase $C$). Suppose that $R$ differs from $R_0$ through the uniform
strain tensor $\epsilon$ and the nuclear displacements $u'$:
\begin{equation}
R = R_0 \cdot (1+\epsilon) + u',
\label{eq:R}
\end{equation}
where all site, vector, and tensor indices are understood, and the dot
product implies a sum over all relevant site and component
indices. Expanding $V$ to second degree in $\epsilon$ and $u'$, we
get:
\begin{eqnarray}
V[R] &= & V[R_0] + \frac{1}{2} u' \cdot \frac{\partial^2 V[R_0]}{\partial
u' \partial u'}\cdot u' \nonumber \\
&& + u' \cdot \frac{\partial^2 V[R_0]}{\partial
u' \partial \epsilon} \cdot \epsilon  + \frac{1}{2} \epsilon \cdot
\frac{\partial^2 V[R_0]}{\partial \epsilon \partial \epsilon} \cdot
\epsilon.
\end{eqnarray}
Making the substitution $u' = u + u_0$ with
\begin{equation}
u_0 = - \left(\frac{\partial^2 V}{\partial u' \partial u'}\right)^{-1}
\cdot \frac{\partial^2 V}{\partial u' \partial \epsilon} \cdot
\epsilon,
\end{equation}
we obtain
\begin{equation}
H = T + \frac{1}{2} u \cdot \frac{\partial^2 V}{\partial u' \partial
u'} \cdot u + \frac{1}{2} \epsilon \cdot {\bf C} \cdot \epsilon,
\end{equation}
where
\begin{equation}
{\bf C} = \frac{\partial^2 V}{\partial \epsilon \partial \epsilon} -
\frac{\partial^2 V}{\partial \epsilon \partial u'} \cdot \left(
\frac{\partial^2 V}{\partial u' \partial u'} \right)^{-1} \cdot
\frac{\partial^2 V}{\partial u' \partial \epsilon}.
\label{eq:C}
\end{equation}
Thus, when $\partial^2 V / \partial u' \partial u'$ becomes soft, that
is, when it has eigenvalues which are nearly zero, ${\bf C}$ will also
become soft according to Eq.(\ref{eq:C}), provided the corresponding
eigenfunctions of ${\partial^2 V}/{\partial u' \partial u'}$ couple to
$\epsilon$.  In materials exhibiting PA, the modes which soften under
pressure will couple to strains associated with elastic moduli that
likewise soften under pressure.\cite{bing92} We can thus focus
primarily on the softening of $\partial^2 V / \partial u' \partial u'$
and treat the strains implicitly, recognizing that the $u$ in Eq.(7)
contains them.

The transformation
\begin{equation}
X_{k\lambda} = \sum_{l\alpha}
\exp{(ikR_l)}S_{\lambda\alpha}(k)u_{l\alpha}
\label{eq:normalmode}
\end{equation}
diagonalizes $H$. In Eq.(\ref{eq:normalmode}), the site index of
Eq.(\ref{eq:kinetic}) has been decomposed into a unit cell index $l$,
with $R_l$ the center of the $l$-th cell, and an internal index
$\alpha$. $\lambda$ is the branch index of the normal mode $k\lambda$,
$k$ its wave vector in the first Brillouin zone, and
$S_{\lambda\alpha}(k)$ is a unitary matrix for each $k$. The squares
of the corresponding normal-mode frequencies $\omega_{k\lambda}$ are
eigenvalues of the dynamical matrix $M^{-1/2}\cdot \partial^2 V /
\partial u' \partial u' \cdot M^{-1/2}$. Thus if $\omega_{k\lambda}$
has a soft, flat branch in which $\omega_{k\lambda}\rightarrow 0$ for
much of the Brillouin zone, so does $\partial^2 V / \partial u'
\partial u'$, as the mass matrix $M$ is positive definite. Besides
that of $\alpha$-quartz under compression, there are other known
examples of soft, flat branches.  In the perovskite ferroelectrics,
for example, there are optical branches which are unstable in the
high-symmetry cubic structure, which, in that case, result in a
ferroelectric phase transition.\cite{ghosez}

We suppose that the subset $\{\mu\}$ of the set $\{\lambda\}$ of all
branches are soft and flat. We construct their associated lattice
Wannier functions\cite{karin} from the corresponding normal modes via
the transformation
\begin{eqnarray}
Y_{l\beta} & = & \frac{1}{N} \sum_{k\lambda \in \{\mu\}}
A_{\beta\lambda}(k) \exp{(-ikR_l)}
X_{k\lambda} \nonumber \\ 
& = & \sum_{m\alpha} S_{l\beta,m\alpha} u_{m\alpha}.
\end{eqnarray}
Since the $S_{l\beta,m\alpha}$ decay exponentially in $|R_m-R_l|$, the
$Y_{l\beta}$ are local modes, optimally localized by proper choice of
the phases and amplitudes of the unitary matrix elements
$A_{\beta\lambda}(k)$; the flatter the branches $\{\mu\}$, the more
localized. In the Einstein limit of zero dispersion, the $Y_{l\beta}$
are confined to a single cell. Also, the Hamiltonian contains no
harmonic coupling between $Y_{l\beta}$ and $X_{k\lambda'}$ for those
$\lambda'$ not contained in $\{\mu\}$.  These local modes form a basis
for the construction of an effective Hamiltonian\cite{karin} much
easier to use for simulations\cite{wag,dudes} and, in our case, much
easier to model than the original Hamiltonian.

\section{The model Hamiltonian}

Ignoring the uniform strain $\epsilon$, the Hamiltonian in
Eq.~(\ref{eq:hamil}) can be decomposed into contributions from the
soft flat branches $\{\mu\}$, the remaining branches $\{\mu'\}$, and
their interaction arising from anharmonicity in $V[R]$:
\begin{equation}
H = H_{\{\mu\}} + H_{\{\mu'\}} + H_{\{\mu\}\{\mu'\}}.
\end{equation}
In developing the model Hamiltonian, only the contribution of the soft, flat
branches, $H_{\{\mu\}}$, is considered explicitly. $H_{\{\mu'\}}$ is replaced
implicitly by the Hamiltonian of a thermal reservoir at temperature $T$ and
$H_{\{\mu\}\{\mu'\}}$ is replaced by an appropriate coupling of the $\{\mu\}$
to the reservoir.

In what follows we drop the subscript $\{\mu\}$ from
$H_{\{\mu\}}$. The dependence of the potential $V$ on the local mode
amplitudes $Y_{l\beta}$ can be expressed as a cluster expansion so
that $H$ takes the form
\begin{eqnarray}
H &= &\sum_{l\beta} \frac{P_{l\beta}^2}{2M_{\beta}} + \sum_l V_l
(Y_{l\beta}) + \sum_{l\ne m} V_{lm}(Y_{l\beta}, Y_{m\beta'})\nonumber \\
& & + \sum_{l\ne m\ne n} V_{lmn}(Y_{l\beta}, Y_{m\beta'}, Y_{n\beta''}) +
{\cal O}(Y^4),
\end{eqnarray}
where $\beta, \beta', \beta'',... \in \{\mu\}$. We now assume all
essential physics remains if we limit the anharmonicity to the local
term $V_l(Y_{l\beta})$, a restriction which simplifies $H$ to
\begin{eqnarray}
H & = & \sum_{l\beta} \frac{P_{l\beta}^2}{2M_{\beta}} + \sum_l
V_l(Y_{l\beta})\nonumber \\
& & + \frac{1}{2} \sum_{l\ne m} V^{\beta\beta'}_{lm}(Y_{l\beta} -
Y_{m\beta})(Y_{l\beta'} - Y_{m\beta'}).
\end{eqnarray}

If we now suppose that only a single branch becomes soft and flat, $H$
further simplifies, after dropping the index $\beta$, to
\begin{equation}
H = \sum_l \frac{P_l^2}{2M} + \sum_l V_l(Y_l) + \frac{1}{2} \sum_{l\ne
m} V_{lm}(Y_l -Y_m)^2.
\end{equation}

As described in the Sec.~II, we are dealing with strongly first-order
phase transitions. The simplest functional dependence of $V_l(Y_l)$
which can lead to such phase transitions is that of a sixth degree
polynomial,
\begin{equation}
V_l = \frac{1}{2} A Y_l^2 - \frac{1}{4} B Y_l^4 + \frac{1}{6} C Y_l^6,
\label{eq:localpot}
\end{equation}
where $B$ and $C$ are positive, and $A$ can be positive or
negative. Not enough is gained by including coefficients $V_{lm}$ beyond nearest
neighbors to warrant the added complexity and increase in the number of
parameters. The harmonic coupling term is then taken to be simply ${1 \over 2}
V_1 \sum_{\langle l,m\rangle} (Y_l - Y_m)^2$, where the notation $\langle l,m
\rangle$ implies that the sum is restricted to nearest-neighbors only.

We now scale displacement by the factor $L = \sqrt{B/C}$, time by $t_0
= B^2/MC$, momentum by $ML/t_0$, and energy by $B^3/C^2$. The
resulting scaled Hamiltonian has the form
\begin{equation}
H = \sum_l \frac{1}{2} P_l^2 + \sum_l V_{\rm loc}(u_l) + \frac{1}{2} k
\sum_{\langle l,m \rangle} (u_l - u_m)^2,
\label{eq:rescaledhamil}
\end{equation}
where $H$ is the scaled Hamiltonian and
$P_l$ and $u_l$ are the scaled local-mode momenta and
amplitudes. $V_{\rm loc}(u_l)$ takes the form
\begin{equation}
V_{\rm loc}(u_l) = \frac{1}{2} \alpha u_l^2 - \frac{1}{4} u_l^4 +
\frac{1}{6} u_l^6,
\end{equation}
where
\begin{equation}
\alpha = \frac{AC}{B^2},
\end{equation}
and $k$ in Eq.~(\ref{eq:rescaledhamil}) is
\begin{equation}
k = \frac{V_1 C}{B^2}.
\end{equation}
For definiteness we take $V_1$, and therefore $k$, to be positive. $H$
thus depends only on the two parameters $\alpha$ and $k$. We take it
to be classical.

The local mode amplitudes $u_l$ evolve with time under the influence
of $H$ and the coupling to the reservoir. We presume $u_l$ to follow
Langevin dynamics with the equation of motion
\begin{equation}
\ddot{u}_l = -\frac{\partial V_{\rm tot}}{\partial u_l} -
\frac{1}{\tau} \dot{u}_l + F_l,
\end{equation}
where
\begin{equation}
V_{\rm tot} = \sum_m V_{\rm loc}(u_m) + \frac{1}{2} k \sum_{\langle
m,n \rangle} (u_m - u_n)^2
\label{eq:Vtot}
\end{equation}
is the total potential. $F_l$ is a Gaussian random force exerted by the
reservoir with moments
\begin{eqnarray}
&& \langle F_l(t) \rangle = 0,\;\mbox{and} \\
&& \langle F_l(t) F_m(t') \rangle = {\cal F} \delta_{lm}\delta(t-t').
\end{eqnarray}
Equipartition requires that
\begin{equation}
\langle \dot{u}_l^2 \rangle = \frac{1}{2} k_B T
\end{equation}
with $T$ the reservoir temperature. The fluctuation-dissipation theorem
requires that
\begin{equation}
\frac{1}{\tau} = \frac{{\cal F}}{k_B T}.
\end{equation}
Thus we are faced with a four-parameter problem $(\alpha, k, \tau, T)$
even with this simplest of models. Achieving a detailed understanding
of the physics of the model requires extensive numerical
simulation. However, many of its important qualitative features are
readily extracted, yielding important insights into the nature of
pressure amorphization in those materials to which the model might be
relevant, as described in the next four sections. The results of the
simulations will be reported separately.\cite{gang}

\section{Pressure amorphization}

We now proceed to eluciate the potential energy landscape of our
model.  For $\alpha > 3/16$ and $\forall k$ in Eq.(\ref{eq:Vtot}),
$V_{\rm tot}$ has a global minimum at $u_l = 0$, $\forall l$, at which
$V_{\rm tot}=0$. We assign this minimum to the low-pressure crystal
structure $C$ described in Sec.~II. When $\alpha < 3/16$ and $\forall
k$, there are two equivalent global minima with
\begin{eqnarray}
&& u_l = w = \left[ \frac{1}{2} + \sqrt{\frac{1}{4} - \alpha}
\right],\;\forall l,\;\mbox{and}
\label{eq:w} \\
&& u_l = -w = -\left[ \frac{1}{2} + \sqrt{\frac{1}{4} - \alpha}
\right],\;\forall l
\end{eqnarray}
at which
\begin{equation}
V_{\rm tot} = \frac{1}{3}N [\frac{3}{4}\alpha - \frac{1}{8} -
(\frac{1}{4}-\alpha)^{3/2}] < 0.
\label{eq:energymin}
\end{equation}
We take this structure to represent the high-pressure crystal
structure $C'$ of Sec.~II. Thus, there is a first-order phase
transition at $\alpha = 3/16$ from $C$ to $C'$ as $\alpha$ decreases
monotonically with pressure $p$ irrespective of the value of $k$. The
two spinodals occur at $\alpha = 1/4$ and $\alpha = 0$, $\forall k$. 
For the phase transition $C\leftrightarrow C'$ to be strongly first-order in
the sense used in Sec.~II., a variation in $\alpha$ from $1/4$ to 0
can be thought of as occuring via substantial change of pressure. The
barrier between $C$ and $C'$ at the transition (where $\alpha = 3/16$)
is 1/96, so that the reduced temperature $T$ must be taken
substantially lower than 1/96 for strong hysteresis to occur, as it does
in type~I materials.

We first consider the Einstein case in which $k = 0$. For $3/16 <
\alpha < 1/4$, $2^N$ degenerate local minima exist in $V_{\rm tot}$
with energy given by Eq.(\ref{eq:energymin}), except that $V_{\rm
tot}$ is positive in this range. The magnitude of $u_l$ is given by
$w$ in Eq.(\ref{eq:w}), its sign being arbitrary. As $\alpha$ falls
below $3/16$, all $2^N$ phases become equivalently stable
relative to $C$. Among these $2^N$ phases, two correspond to {\it
single-crystal} $C'$, and an exponentially small fraction to {\it
poly}- or {\it micro-crystalline} $C'$; yet the vast majority of the
structures are {\it amorphous} in that there is at most short-range
order in the $u_l$ even though a clear topological memory of the
original crystal structure $C$ remains. Barriers keep the system in
$C$ until the spinodal at $\alpha = 0$ is approached. At the spinodal,
$C$ becomes locally unstable relative to all $2^N$ phases. The
probability of the system finding its way from $C$ to one of the
amorphous phases is vastly higher than that to one of the
exponentially fewer polycrystalline $C'$ phases.

More generally, pressure amorphization occurs if (i) there is a region
of the $\alpha$-$k$ plane in which the $a$ phases are more stable than
the $C$ phase but metastable with respect to $C'$ and (ii) if $\tau$
and $T$ are such that $a$ is preferentially accessible from $C$ either
irreversibly (types~I, II, and~IV) or temporarily (types~III
and~V). In the next three sections we shall argue that this indeed
holds for the 2-site case, the 3-site case, and the $N$-site case by
examining the topology of the potential-energy landscape presented by
$V[u]$. In reading these sections, it is important to bear in mind
that $k$ determines how flat the branch is and $\alpha$ how soft. A
soft, flat branch has small $\alpha$ and small $k$.

Note that the previous discussion straightforwardly applies to the
case in which there are $\nu$ flat branches. $V[Y]$ is then defined on
a $\nu N$-dimensional configuration space. Around every minimum in
this space there is a basin of attraction bounded by a separatrix, a
hypersurface of dimension $\nu N - 1$. As in the single-branch case,
PA will occur when the amorphous minima $a$ are preferentiably
accessible from $C$.

\section{The 2-site model}

To begin, we study pressure amorphization in our simple Hamiltonian
through analysis of the topology and kinetics of a 2-site model.  In
what follows the local-mode amplitude and momentum of the first site
are $x$ and $P_x$, those of the second site $y$ and $P_y$. The
Hamiltonian is
\begin{equation}
H = T + V,
\end{equation}
where
\begin{equation}
T = \frac{1}{2}(P_x^2 + P_y^2)
\label{eq:twositekin}
\end{equation}
and
\begin{eqnarray}
V & = & \frac{1}{2}\alpha(x^2+y^2) - \frac{1}{4} (x^4+y^4) + \frac{1}{6}
(x^6 + y^6)\nonumber \\ && + \frac{1}{2}k(x-y)^2.
\label{eq:2sitepot}
\end{eqnarray}	
In Fig.~1, we show in the $xy$ configuration plane the critical points
of $V(x,y)$ for $k=0$ and $0<\alpha<3/16$. $C$ occurs at $x=y=0$, $C'$
at $x=y=\pm w$. There is only one saddle point separating the $C$,
$C'$, or $a$ basins from each neighboring intermediate $i$ basin.

\begin{figure}[t!]
\centering
\includegraphics[width=\columnwidth]{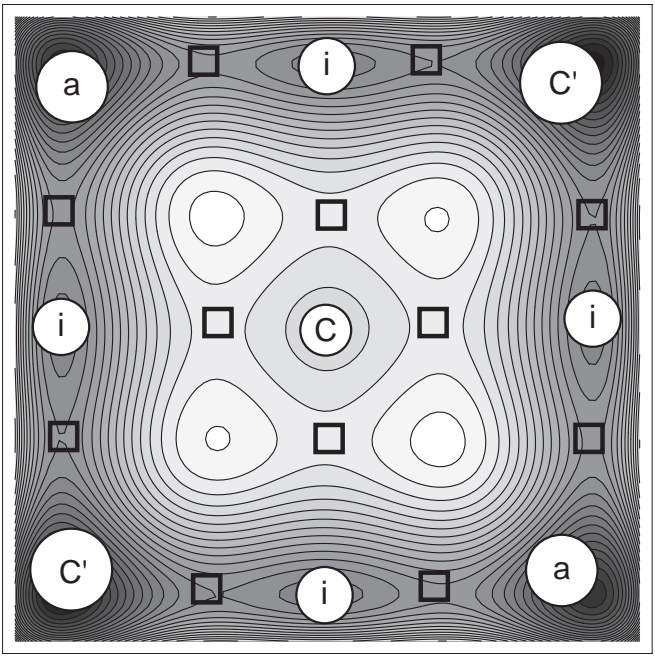}
\caption{Contour plot of the potential $V(x,y)$ of the 2-site model in
the $xy$ plane. Minima are indicated by white circles and saddle
points by open squares; the maxima are clear by inspection.  The
larger the symbol, the lower the value of $V(x,y)$.  The crystalline
minima are indicated by $C$ or $C'$, the intermediate minima by $i$,
and the amorphous minima by $a$.}
\label{fig:1}
\end{figure}

In Fig.~2, we divide the $\alpha$-$k$ phase plane into regions each of
which has a distinct critical point (CP) structure, labelling each
region (with one exception) by the stable crystal structure within it
and displaying the CP structure. The exception, shaded, is the region
within which $V$ retains at finite $k$ for $0<\alpha<3/16$ the same
topology, i.e. CP structure, that it has for $k=0$ as shown in
Fig.~1. It is only within this region that PA is possible because the
$a$ minima exist, they are stable relative to $C$, and both $a$ and
$C'$ can be reached from $C$ only by trajectories which must pass
through intermediate basins $i$ so that the transition $C$ to $a$ can
compete kinetically with $C$ to $C'$.

\begin{figure}[t!]
\begin{center}
\includegraphics[width=\columnwidth]{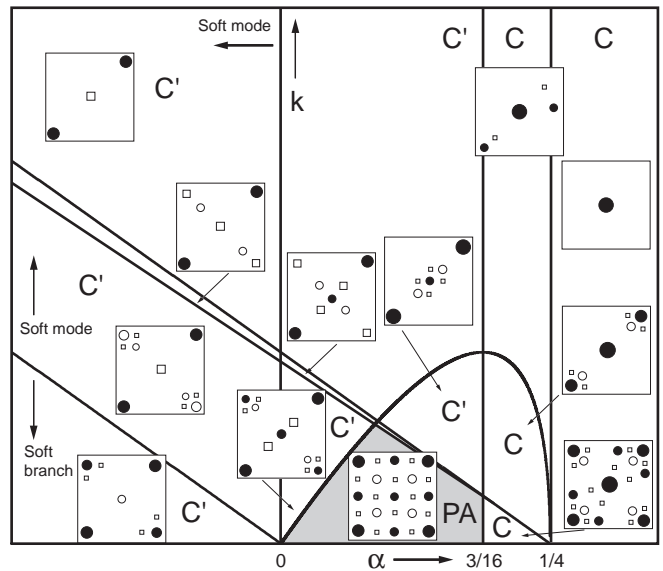}
\end{center}
\caption{The $\alpha$-$k$ phase plane of the 2-site model divided into
regions of distinct critical point structure.  Maxima are indicated by
open circles, minima by solid circles, and saddle points by open
squares.  The sizes of the circles reflect the relative magnitudes of
the extrema.  Each region is labelled by the stable crystal structure
within it, except for one. Within that region, marked PA, the critical
point structure is the same as for $k=0$, $0<\alpha<3/16$, and
pressure amorphization is possible as discussed in the text.}
\label{fig:2}
\end{figure}

For PA to occur within this 2-site model, trajectories from $C$ to $a$
must be traversed with some probability relative to those from $C$ to
$C'$.  This happens if the following seven conditions are
met:\cite{foot1} (i) The minimum must exist and be stable relative to
$C$; (ii) Intermediate basins of attraction must exist, and all
statistically significant trajectories from $C$ to $C'$ or $a$ must
pass through them; (iii) For hysteresis, both $V_{ic}-V_c$ and
$V_{ia}-V_a$ must be much larger than $k_B T$ at $\alpha = 3/16$; (iv)
For $ia$ to be accessibile from $ic$, one requires $V_{ic}-V_{ia}
\gtrsim 0$ and slow relaxation in the $i$ basin; (v) For the
probabilities of transitions from $i$ to $a$ and from $i$ to $C'$ to
be of comparable magnitudes, $V_{ia}-V_{ic'}$ must not be much larger
than $k_BT$; (vi) For the transition from $i$ to $a$ to be
irreversible, or likewise for there to be a very long resident time in
$a$ before a transition from $a$ to $C'$, $V_{ia}-V_a \gg k_BT$ and
fast relaxation must occur in the basin; (vii) Finally, for
consistency of (v) and (vi), $\tau$ must not be too low. Detailed
qualitative considerations strongly suggest that these conditions can
be met only for $\alpha$ and $k$ in the shaded region of Fig.~2,
provided that the remaining parameters $T$ and $\tau$ assume
appropriate values. The actual determination of the probability of
$C\rightarrow a$ relative to that for $C\rightarrow C'$ for various
values of $T$ and $\tau$ requires numerical simulation, the results of
which will be reported elsewhere.

We have supposed that $\alpha$ is a monotonically decreasing function
of pressure. For simplicity, we shall further suppose that $k$ depends
weakly on $p$ as $\alpha$ decreases. The effect of pressure change on
the system can be represented by a trajectory in the $\alpha$-$k$
plane. One can readily see that the five types of behavior described
in Sec.~II have direct parallels in the
responses of our model system along different trajectories in the
$\alpha$-$k$ plane. For example, a trajectory which starts at
$\alpha>1/4$ and moves well into the PA region before reversing and
returning corresponds to type~I behavior, with $p_{ca}$ corresponding
to $0<\alpha_{ca}<3/16$ and $p_{ac}$ corresponding to $3/16<
\alpha_{ac}<1/4$. Type~II corresponds to a trajectory that starts in
structure $C$ at $\alpha=\alpha_0$ with $3/16 < \alpha_0 < 1/4$, transforms into $a$ below
$0< \alpha_{ac} < 3/16$, and can return to $\alpha_0$ while still
being amorphous. Trajectories terminating in the PA region near its
upper right boundary correspond to behavior typical of type~III
materials. A trajectory which starts at $\alpha>1/4$ and passes
through the PA region into that with $\alpha < 0$ would correspond to
the situation in a type~IV material. Finally, trajectories which, on
their return to larger $\alpha$, pass through the upper right boundary
can undergo an $a\rightarrow C'$ transition inside the boundary,
remaining in $C'$ upon pressure reduction until some $\alpha$ in
$(3/16,1/4)$ is reached; these would correspond to transitions in
type~V materials.

\section{The 3-site model}

We now discuss an extension of the model described in Sec.~VI to three
sites.  We label the third site by $z$ and suppose it is coupled only
to $y$ as in a linear triatomic molecule. An additional term $P_z^2/2$
is added to $T$ in Eq.(\ref{eq:twositekin}), and $V$ becomes
\begin{eqnarray}
&&V = \frac{1}{2}\alpha(x^2+y^2+z^2) - \frac{1}{4} (x^4+y^4+z^4)
\nonumber \\ & & +
\frac{1}{6} (x^6 + y^6+z^6) + \frac{1}{2}k[(x-y)^2 + (y-z)^2].
\end{eqnarray}
Just as in the 2-site model, there is a region in the $\alpha$-$k$
plane within which the CP structure and the topology of $V$ remain the
same as for $k=0$. It is within this region that PA can occur. Its
shape is qualitatively the same as that for the 2-site model, bounded
by the $\alpha$-axis from 0 to $3/16$ and on the right by a vertical
segment at $\alpha = 3/16$. The upper boundary segments, however, are
displaced towards smaller $k$.

\begin{figure}[t!]
\begin{center}
\includegraphics[width=\columnwidth]{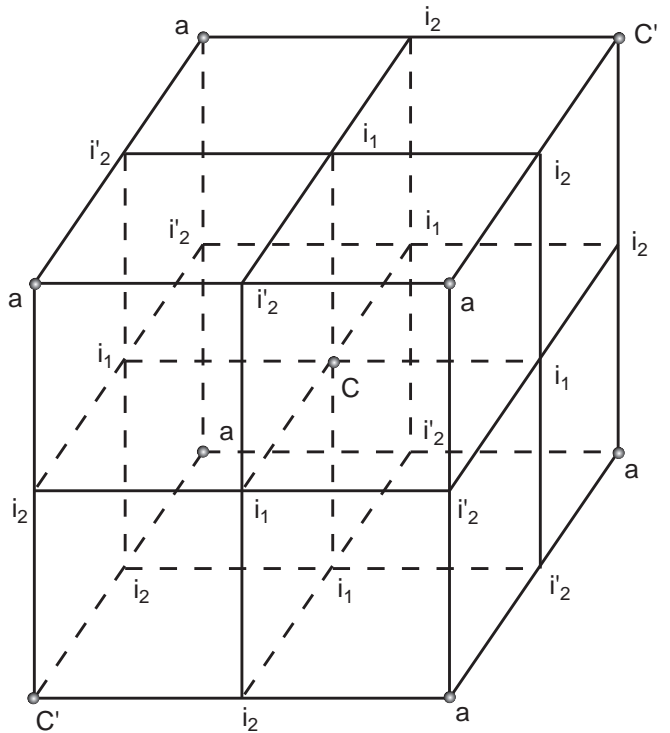}
\end{center}
\caption{Minima of the potential energy landscape of the 3-site
model.}
\label{fig:3}
\end{figure}

The $k=0$ minima are shown in Fig.~3. For small but finite $k$ in the PA
region, the structure is compressed perpendicular to the $x = y$ and
$y = z$ planes. There are now six classes of minima: one $C$, six
$i_1$, six $i_2$, six $i_2'$, six $a$, and two $C'$, using the
notation of Fig.~3. The sequence of values of $V$ at the minima is
$V_C > V_{i_1} > V_{i'_2} > V_{i_2} > V_a > V_{C'}$. The sequence at
the saddles is $V_{ci} > V_{i_1 i_2'} > V_{i_1 i_2} > V_{i_2 a},\,
V_{i_2' a},\, V_{i_2'C'}$ and $V_{ci} - V_{i_1 i_2'} \gg V_{i_1 i_2'}
- V_{i_1 i_2}$, etc. These sequences have the following
implications. As before, transitions between the basins occur over
separatrices near the transition states (saddle points). Relaxation
can be fast enough to bias the transformation down hill but still slow
enough that enough energy remains to get over each successive barrier
with little difference between rates over the primed and unprimed
barriers. To reach $C'$ from $C$, the system must follow the sequence
$C\rightarrow i\rightarrow i_2 \rightarrow C'$ for which there are 12
independent routes. To reach $a$ from $C$, the system can follow
either of two sequences, $C\rightarrow i \rightarrow i_2 \rightarrow
a$ or $C\rightarrow i \rightarrow i_2' \rightarrow a$, for which there
are 36 routes, or 3 times as many as for $C\rightarrow C'$. This
increase in the number of paths from $C$ to $a$ relative to that from
$C$ to $C'$ compensates for the more favorable kinetics of the final
step from $i_2$ to $C'$, increasing the probability of $C\rightarrow
a$ relative to that of $C\rightarrow C'$ over that of the two-site model.

\section{The $N$-site model}

We now jump to a model with a macroscopically large number of
equivalent sites each of which is coupled to a finite number of
nearest neighbors via the harmonic term in $V$. Once again, there is a
domain in the $\alpha$-$k$ plane in which the topology of $V$ is the
same for finite $k$ as for $k=0$. Its lower boundary is unchanged at
$k=0$, $0<\alpha<3/16$. Its upper boundary is further compressed, the
degree of compression increasing with the number of nearest
neighbors. Just as the minima for the 3-site case lie at the center,
face centers, edge centers, and corners of a cube, so are the minima
arranged on an $N$-dimensional hypercube in the $N$-dimensional
configuration space for $k=0$. As $k$ increases, the CP structure
distorts without change of topology up to some $k$ which depends on
$\alpha$. The existence of this finite domain of topological stability
is guaranteed by the analyticity of $V$, its gradient, and its Hessian
in the configuration space. There is now one minimum of type~$C$,
$3^N-2^N-1$ of type~$i$, $2^N-2$ of type~$a$, and 2 of type~$C'$. The
sequence of values of $V$ is $V_c > V_i > V_a >V_{C'}$.\cite{foot2}
The number of $a$ minima is exponentially larger than the number of
$C'$ minima, even if one extracts the polycrystalline $C'$ structures
from the above number of $a$'s and adds them to the $C'$'s, as one can
see from the following estimate. Let $n$ be the minimum grain
size. Then $2^{N/n}$ is a lower bound but still a good estimate of the
number of polycrystalline structures. The corrected ratio of $C'$'s to
$a$'s is thus estimated to be
%
\begin{equation}
\frac{2^{-N(1-1/n)}}{1-2^{-N(1-1/n)}} \xrightarrow[N\rightarrow\infty]{}
 0,\; \mbox{with}\; n > 1,
\end{equation}
which is still exponentially small. The number of paths from $C$ to
$a$ is correspondingly exponentially larger than that from $C$ to
$C'$, overcoming any kinetic advantage of any single path from $C$ to
$C'$. In conclusion, the key to PA within this model is the existence
of a region of the $\alpha$-$k$ plane within which the $k=0$ topology
persists, a region of a flat (small $k$) and soft but not unstable
($0<\alpha<3/16$) branches together with $\tau$ and $T$ in a suitable
range.

\section{Relation to conventional nucleation and growth mechanism}

Suppose the transformation $C\rightarrow C'$ were to occur by the
conventional nucleation and growth mechanism with $\dot{I}$ the
nucleation rate and $\dot{R}$ the growth rate.\cite{turn2} For
$\dot{I}/\dot{R}$ finite, a polycrystal of $C'$ results. In the limit
$\dot{I}/\dot{R} \rightarrow 0$, a single crystal results. In the
limit $\dot{I}/\dot{R} \rightarrow \infty$, a nanocrystalline material
results, and we can imagine that in the ultimate limit in which
nucleation totally dominates growth, an amorphous phase results when
the crystallite is so small that there is no distinction between bulk
and grain boundary. In the classical theory of nucleation, the
critical radius of the nucleus of $C'$ is
\begin{equation}
R^* = \frac{2\sigma}{\Delta g}.
\label{eq:nucleus}
\end{equation}
Here $\Delta g$ is the bulk Gibbs free-energy difference between $C$
and $C'$, and $\sigma$ is the interfacial free energy. The
corresponding barrier to nucleation is
\begin{equation}
\Delta G^* = \frac{16\pi}{3} \frac{\sigma^3}{(\Delta g)^2}.
\label{eq:barrier}
\end{equation}
For nucleation completely to dominate growth and for the critical size
nucleus to be so small that the result is an amorphous phase, it is
obvious from Eqs.~(\ref{eq:nucleus}) and~(\ref{eq:barrier}) that
$\sigma$ must become small. $\sigma$ measures the free-energy cost of
a rapid local structural change from $C$ to $C'$. In our simple model,
the energy cost of such a structural change is simply proportional to
the coupling constant $k$. Thus, our finding the possibility of
pressure amorphization in the region of the $\alpha$-$k$ plane within
which the $k=0$ topology of $V$ is preserved is completely consistent
with nucleation theory, the smallness of $k$ corresponding to small
$\sigma$ and the existence of intermediate minima corresponding to
finite $\Delta g$ and slow growth rate. To conclude, PA occurs when
nucleation totally dominates growth.

\section{Summary and conclusions}

In this work, we make a clear distinction between two types of
amorphous phases. The first, the conventional glassy phase, is not
topologically equivalent to any crystalline phase, and glass formation
from the crystal is a reconstructive transformation.\cite{turn2} Such
glasses are formed, for example, by sufficiently rapid cooling from
the melt. A conventional glass can form via pressure amorphization if
(i) the limit of local stability of $C$ relative to its melt (the
spinodal) is approached, and (ii) if the glass transition temperature
of the melt is then higher than the equilibrium melting temperature of
$C$.\cite{pony,scio} Here we are concerned not with conventional
glasses, but with pressure-induced amorphous phases in which the
structural instability that leads to the amorphization is associated
with one or more soft, flat branches of the phonon spectrum and in
which memory of the original topology of the low-pressure crystal
structure is retained through the remaining stable branches. In such
cases, the amorphization arises from a random displacive or
orientational transition instead of from a reconstructive transition.

This type of pressure amorphization is driven by an underlying
strongly first-order structural phase transition $C\rightarrow C'$
under compression. In our view, as $C$ becomes thermodynamically
unstable relative to $C'$, there is one or more branches which flatten
with increasing pressure. Local modes in the form of lattice Wannier
functions provide a natural means of description of such flat
branches. We have supposed that the nonlinearities associated with the
first-order phase transition arise primarily in the amplitudes of the
individual local modes, an inessential simplification. We have argued
that there is a range of pressures within which the local
nonlinearities and relative flatness of the relevant branch or
branches allow an exponentially large number of amorphous structures
to be metastable and preferentially kinetically accessible over $C$
relative to the thermodynamically stable single crystal or
polycrystalline $C'$ phase. Such metastability and kinetic preference
occur within a restricted region of the parameter space defining the
potential energy landscape in the configuration space of the local
modes, as shown explicitly for the 2-site model in Fig.~2. The ambient
temperature must be low enough to lead to substantial hysteresis
($T\ll 1/96$), so that $C$ persists well into its domain of
metastability as $p$ increases. Once the structural transition is
initiated, it proceeds downhill via an exponentially large number of
paths through the basins of attraction of intermediate structures
without being trapped there, requiring relaxation rates in a suitable
range. If trapping were to occur, it would correspond to intermixed
amorphous $a$, intermediate, and crystalline $C$ regions, resulting in
observable Bragg peaks associated with $C$. Further compression (i.e.,
further reduction of $\alpha$) would result in a drop in the fraction
of $C$, decreasing Bragg peak intensities. That is, trapping implies a
broadened $C$ to $a$ transition.  The intermediate basins correspond
to a random admixture of both $C$-like and $C'$-like local
configurations.

Normally, first-order structural transitions are discussed in terms of
nucleation and growth. In the present picture, the critical size
nucleus is not much more than a single unit cell (with minor
displacements in neighboring cells) and nucleation overwhelms growth,
leaving the system in an amorphous structure when nucleation is
complete.

The type of pattern observed in the structural transitions caused by
pressure changes depends on the particular trajectory through the
parameter space of the potential which is induced by the changing
pressure.

The picture developed in this paper can embrace all of the phenomena
which have been observed in association with pressure amorphization.
Nevertheless, the discussion has been largely qualitative. Obtaining a
quantitative demonstration that these simple models admit pressure
amorphization requires numerical simulation.\cite{gang} Once that has
been demonstrated, such simulations should be carried out for real
materials since substantial experience has already been built up in
constructing effective Hamiltonians of the type considered here for
specific substances.\cite{karin,wag,dudes} Among the tasks for
simulations is to distinguish whether the occurence of pressure
amorphization results in a conventional glass or leaves the
crystalline topology invariant for a particular material.

\acknowledgements

We greatly benifitted from discussions with A. Angel, R. Car,
J. R. Chelikowsky, P. Giannozzi, G. N. Greaves, G. S. Grest,
K. M. Rabe, S. Scandolo, and D. Vanderbilt. J. I. acknowledges the
financial suppor of the ONR through Grant N0014-97-1-0048.


\begin{references}
\bibitem{foot1} In the following, single subscripts on $V$ indicate
the value of $V$ at a minimum. Double subscripts indicate its value at
a saddle point lying on the separatrix between the two minima
specified by the subscript.
\end{references}
\end{document}